\documentstyle[11pt]{article}

\def\be{\begin{equation}}
\def\ee{\end{equation}}

\begin{document}

\begin{flushright}
BRX TH-464
\end{flushright}

\begin{center}
{\Large\bf Infinities in Quantum Gravities}

\vspace{.2in}

\renewcommand{\thefootnote}{\fnsymbol{footnote}}
\setcounter{footnote}{0}

S. Deser\footnote{\tt deser@brandeis.edu}

\vspace{.05in}

Department of Physics, Brandeis University, Waltham, MA 02454, USA
\end{center}

\begin{abstract}
We first present a review, intended for classical
relativists, of the ultraviolet difficulties faced
 by local quantum gravity theories
in both the usual Einstein versions and in their supergravity
extensions, at least perturbatively.  These problems, present in
arbitrary dimensions, are traceable to the dimensionality of the
Einstein constant.  We then summarize very recent results about
supergravity at  the highest allowed dimension, D=11, showing that
also this unique model suffers from infinities already at 2 loops,
despite its high degree of supersymmetry.  The conclusion is that
there is no viable nonghost  quantum field model that includes
general relativity. \end{abstract}

\section{Introduction}


This report on the current state of quantum gravity (QG) will have
both a narrower and a broader scope than ``QG": on the one hand, I
will only be concerned with ultraviolet behavior -- the
``renormalizability problem"; on the other,  QG here signifies not
just D=4 quantized Einstein theory (QGR) but more importantly its
extensions to supergravities (SUGRA) in all possible dimensions.
Indeed, my lecture is primarily dedicated to the very new results
I will report on the hitherto unresolved renormalizability
properties of the ultimate such theory, D=11 SUGRA.  This will
lead to the conclusion that there are no viable (but otherwise
physically acceptable) local quantum field theories that contain
or reduce to GR, at least in the perturbative framework available
to us. This conclusion of course in no way negates the validity of
say GR as an effective, tree level, model; it means that there
must be an underlying theory, such as strings, or M-theory, with
extended rather than pointlike excitations. Indeed, the fact that
superstring theories are both finite and reduce to GR in the local
limit is one of their essential attributes.

I will first rapidly review the relevant aspects of QGR, the
quantized Einstein theory, then remind you of the now standard
SUGRA extensions and their most relevant property, namely a new
local invariance -- supersymmetry (SUSY).  We will then turn to a
brief discussion of the renormalizability problem, its fundamental
origin in the dimensionality of the Einstein coupling constant,
and to the ``struggle" between the constraints imposed by higher
symmetries and this dimensionality in QGR (at any D) as well as in
SUGRAs with D$<$11. Finally, we will consider the maximal, and
ultimate, D=11 SUGRA whose newly uncovered nonrenormalizability
properties will be exhibited.  For reasons of space, extensive
referencing cannot be provided nor is this review always
technically precise.

\section{QGR}

Classical GR, well defined in arbitrary dimension D$\geq$4, does
not exist at D=2, and is defined but not dynamical for D=3. Let me
immediately state what models are unphysical already classically
and hence are not considered here: these are theories for which
unitarity would be violated because their spectrum contains ghost
or tachyon excitations or which do not limit to GR at all.  In
particular I exclude both ``R+R$^2$" theories that involve higher
derivative already at the kinetic level (see below) and ``R$^2$"
theories such as Weyl gravity, where there is no Einstein term at
all. Theories with R$^3$ or higher additions to GR share the fate
of QGR, so we can neglect them as well.  [Here ``R$^2$" and
``R$^3$" mean generic powers of Riemann, Weyl, Ricci or scalar
curvatures.] We will also not discuss models with non-vanishing
cosmological constant; they can be quantized but are no better
behaved.

To define QGR, we must choose a vacuum state about which to expand
the metric, the natural choice being flat space, with Minkowski
metric $\eta_{\mu\nu}$. While one might perhaps hope that choosing
different vacua could affect the result, this seems highly
unlikely because the main issue is really the Einstein constant's
dimensionality. Likewise, we do not consider in this context --
because none are known -- any nonperturbative approaches to the
QGR problem. The expansion we will use is
\begin{equation}
g_{\mu\nu} \equiv \eta_{\mu\nu} + \kappa h_{\mu\nu} \; , \;\;\;\;
g^{\mu\nu} = \eta^{\mu\nu} - \kappa \, \eta^{\mu\nu} + .. \; ,
\;\;\;\; h^{\mu\nu} \equiv h_{\alpha\beta}\eta^{\mu\alpha}
\eta^{\nu\beta} %
\end{equation} %
but again field redefinitions, {\it
e.g.}, expanding $\sqrt{-g} \: g^{\mu\nu} \equiv \eta^{\mu\nu} +
H^{\mu\nu}$ fare no better. The Einstein action $I_E = \kappa^{-2}
\int d^Dx \; \sqrt{-g} \: R $ consists of a free part
corresponding to a massless spin 2 field, along with an infinite
series of ever higher order vertices. The propagator term is
essentially
\begin{equation}
I_F = - \textstyle{\frac{1}{2}} \int d^Dx \: (\partial_\alpha
h_{\mu\nu} )^2 \; .%
\end{equation}%
 This expression shows why
adding $\int d^Dx \, R^2$ terms is forbidden: their propagator
contributions are $\sim \int d^Dx \, (\partial h) \partial\partial
(\partial h)$, thereby introducing ghost/tachyon states that
violate unitarity and energy stability, since the propagator is
now a sum of two excitation modes, $\sim p^{-2} - (p^2 + \alpha
)^{-1}$ with relative ghost sign, tachyonic mass, or both. It
should also be emphasized that, at QGR level, any R$^2$ terms that
arise as radiative (loop) corrections to an underlying purely GR
action are {\it not} to be used to modify the propagator part (2);
by the perturbation-theoretic rules, they can only act as vertex
corrections, without altering the $p^{-2}$ behavior of the
propagator. The infinite series of vertex terms,
\begin{equation}
I_E - I_F \sim \sum^\infty_{n=1} \int d^Dx  (\partial h)^2 (\kappa
h)^n \end{equation}
is characterized by the rising power in
$\kappa$ and a common 2-derivative character.  Quantization turns
$h_{\mu\nu}$ into an operator, and in units $\hbar = 1=c$,
$\kappa^2$ has dimension $[L]^{D-2}$; it is dimensionless only at
D=2, where the action is just the integrated topological Euler
density, and so is empty.

\section{SUGRA}

Before proceeding to the UV problem, we quickly define the basics
of SUGRA.  These theories are fermionic extensions of QGR that
include at least a spin 3/2 massless field $\psi_\mu$ and
(depending on the dimension) perhaps also other $(3/2 \geq s \geq
0))$ lower spin ``matter". A SUGRA is thus simply ``just another"
GR + matter system, but with two deep differences: the most
important is the presence of a new local SUSY Grassmanian gauge
invariance that interchanges bosons with fermions and constrains
its form very strongly; second, because fermions are involved and
those are only consistent when (second-) quantized, their gravity
and other bosonic components must also be quantized {\it ab
initio}; there is not classical SUGRA (though it does have a
classical bosonic limit).  For concreteness, I remind you (but
only schematically) of the original D=4 N=1 SUGRA \cite{ferrara},
\begin{equation}
I_{SG4} = \int d^4x \left[ \kappa^{-2} \sqrt{g} \, R - i/2 \:
\bar{\psi}_\mu \; /\,\!\!\!\!\!D \psi^\mu \right] \end{equation}
where $D\!\!\!\! /$ is the (covariant) Dirac operator; the SUSY
here is
\begin{equation}
\delta e_{\mu a} = \bar{\alpha} (x) \gamma_a \, \psi_\mu \; ,
\;\;\; \delta \psi_\mu = D_\mu \, \alpha (x) \end{equation} with
$\alpha (x)$ is a Grassmann (anticommuting) function and the
graviton is ``hidden" in the covariant derivative of the spinor
$\alpha (x)$. Another fundamental difference between SUGRAs and
GR+matter is that the former must satisfy conditions such as
equality of numbers of bose/fermi excitations in order for the
SUSY algebra to close.  This not only restricts the permitted
SUGRAs at any dimension D, but also provide an upper bound, D=11,
on the dimension at which SUGRA can exist: beyond D=11, massless
spins higher than 2 are required (notoriously difficult to couple
consistently to gravity) and there will be more than one graviton
as well.  We will discuss D=11 and its special properties later.

\section{Nonrenormalizability in Gravitational Theories: QGR}

Let us begin for contrast with the ultraviolet problem in say
QED$_4$, where the coupling constant $\alpha$, which also counts
closed loop orders (after the usual eA$\rightarrow$A rescaling),
is dimensionless. The ultraviolet (UV) problem -- when calculating
closed loop corrections to any process, there is in general an
infinity -- is exemplified by the one-loop correction to the
electron's mass due to a virtual photon line, that contains a
logarithmic divergence proportional to the electron mass term
itself. There are also (essential!) nonlocal, finite, terms, but
this local term can (and must) be removed at the price of
introducing the indeterminacy of the bare mass $m_0 \rightarrow
m_0 + \delta m$ into the theory. Fortunately, here as well as to
any $n$-loop  $(\sim \alpha^n )$ order, the counterterm in
question retains the {\it form} of the original QED action's mass
term since there is no new dimensional quantity in the problem.
Thus, all infinities can be removed at the cost of redefining a
{\it finite} number of parameters (here charge and mass) in the
theory.  This is all very familiar; what about dimensionfull
couplings? Already in the thirties, Heisenberg noted that theories
with coupling constants (and hence loop expansions) having
positive dimension (for him the Fermi 4-fermion weak interaction
model $(\sim G \int d^4x (\bar{\psi} \psi )^2 )$) would have
increasing divergences at each order, or in modern language
require an ever-increasing number of counterterms and hence of
undetermined parameters in the theory: one would need to specify
an infinite, and not just a finite, set of parameters before the
theory is fixed. [Parenthetically, there are other problems that
make QED$_4$ alone unsatisfactory; by contrast Yang--Mills (YM)
theory, while also renormalizable, has other advantages
(asymptotic freedom) that effectively obviate the whole
counterterm apparatus, and there are even theories with negative
dimensional couplings that are ``superenormalizable". But all that
is irrelevant here. Let me finally allude to one conceivable
nonperturbative ``solution" of the divergence problem.  Many years
ago, Weinberg suggested that models might exist that are
perturbatively unacceptable, but contain fixed points that would
imply well-defined predictions when viewed nonperturbatively.  To
my knowledge, there is as yet no realistic application of this
idea to gravity.]

What loopholes are left by the above generic linkage between
dimensional coupling and nonrenormalizability? There are basically
two. First, the infinite counterterms may all be present but
harmless, because they are in some way absorbable without having
to introduce a new parameter; second, they may actually not arise
at all because forbidden on invariance grounds. [Of course, there
could also be accidental cancellations of infinities, but an
infinite number of them is infinitely unlikely; we know that (also
in physics) what is not forbidden is generically compulsory.]  We
will see examples of both loopholes, but not for our physical QG
models.

Consider first the classic example of source-free D=4 QGR, which
was also the first to be studied systematically \cite{thooft}.
Here $\kappa^2 = [L]^2$, so the candidate infinite counterterms
are essentially
\begin{equation}
\Delta I_{4} =``\infty " \sum^\infty_{n=0} \int d^4x (\kappa^2
R)^n \: R^2 \end{equation} that is, the one-loop candidate
infinity is $\sim \kappa^0 \int R^2$, the 2-loop one is $\kappa^2
\int R^3$ etc.  Now explicit calculation shows that there is no
miracle -- the one-loop term is there, but it does benefit from
the first of our loopholes, in a way that I merely indicate here:
If a counterterm is of the form $\int \delta I_0/ \delta \chi \:
X$, where $\delta I_0/\delta \chi = 0$ is the original system's
field equation (here $R_{\mu\nu} =0$) and $X$ is any quantity,
then this term can be absorbed by a field redefinition, $\chi
\rightarrow \chi + ``\infty " \, X$, with no renormalization
required \cite{thooft}. But, thanks to the D=4 Gauss--Bonnet
theorem, $\int d^4x \, R^2_{\mu\nu\alpha\beta} = \int d^4x (4
R^2_{\mu\nu} - R^2 )$, all possible ``$R^2$" terms can be written
as $\int d^4x \: R_{\mu\nu} X^{\mu\nu}$ and hence removed by
redefining the metric, and the theory is one-loop safe. What about
2 loops? This is a more difficult matter, involving a very nasty
calculation of the coefficient of the on-shell non-vanishing local
$\int d^4x$ (Weyl)$^3$ term; two heroic calculations \cite{goroff}
indeed established its presence. Of course, we will never strictly
know whether the infinities stop after some finite loop number,
but it is safe to say the theory is not viable (barring a truly
surprising as yet unknown and infinitely strong symmetry of the
Einstein action).  The same story holds for QGR in any D$>$4,
since it contains D=4 as a dimensional reduction and the reduction
cannot make a finite theory infinite.  In any case, pure QGR is
rather academic, as already at 1-loop even D=4 QGR plus any matter
of spin 0, $\frac{1}{2}$ or 1 is bad, \cite{thooft,sdvn} because
the infinities do not just enter in the form $``\infty " \int
(G^{\mu\nu} - T^{\mu\nu} ) X_{\mu\nu}$ and so are not redefinable
away.

\section{D$<$11 SUGRA Infinities}

What about spin 3/2 ``matter"?  After all D=4 SUGRA can be thought
of in that way as well, so one would expect a one-loop
catastrophe. However, there {\it is} a new symmetry at play here,
namely the SUSY of (5).  Indeed, this forces the counterterms to
be no worse than for pure QGR: they are, as for pure QGR,
proportional to the {\it full} (SUGRA) equations of motion.  So
far, this merely makes SUGRA no worse than QGR without sources.
But in fact, it {\it is} better as a result of our second,
``higher symmetry", loophole at 2 loops: there exists no SUSY
completion of the $R^3_{{\mu\nu}\alpha\beta}$ term that destroyed
QGR! Does SUSY always ``beat" the curse of $\kappa^2 = [L]^2$?
Alas, no.  At 3 loops \cite{sdjkks}, there is a SUSY invariant,
which happens to start (for deep D=4 and SUSY reasons), as
$\kappa^4 \int B^2_{{\mu\nu}\alpha\beta}$ where $B$ is the
(quadratic in curvature) Bel--Robinson tensor.  The same is true
for extended versions of SUGRA that involve additional fields and
a further symmetry, labelled by an internal index $N$, {\it eg.},
$N$=2 \cite{sdjk};  the series extends all the way to maximal,
$N$=8 SUGRA, but N=8 enjoys a special, hidden, symmetry. One might
suppose that calculation of these coefficients here would be
technically overwhelming, and so it remained until quite recently
with the development of a powerful new formalism \cite{bern}, to
which we shall return. Indeed, it was possible to show that, due
to this symmetry, at least maximal $N$=8 SUGRA enjoys a reprieve
until 5 loops \cite{bern},\cite{howe}. I must refer to the
original papers for the technology that made this possible, but
mention in passing that it relies on a separately interesting
insight, the KLT \cite{kawai} correspondence between closed and
``squares" of open strings that essentially relates graviton
scattering amplitudes to ``squares" of Yang--Mills amplitudes.

So D=4 SUGRAs become infinite as well, at the lowest possible
order consistent with their symmetries; they just start a bit
later in loop order, but there are no accidental cancellations. In
this connection, I should also mention the amusing case of D=3 QGR
where $\kappa^2 \sim [L]^{+1}$ but where there should be no true
infinities simply because the theory has no local excitations.
This comes about because in D=3 Riemann and Ricci tensors are
equivalent so all invariants (not just the one-loop one) vanish on
Einstein shell.  This simple argument is borne out by explicit
calculations in different formalisms \cite{witten}; the roles of
(and insensitivity to) different background choices are nicely
illustrated here.

The new methods I mentioned, together with a reasonable
understanding of how to express SUSY invariants in higher
dimensional SUGRAs, can be used to show that infinities also
appear in all SUGRAs with D$\leq$10, which have their roots in
(D=10) superstring theories. This left, as the only model
outstanding, the last possible, (D=11) SUGRA, to which we now
come.

\section{D=11 SUGRA}

Shortly after the original $N$=1 D=4 SUGRA was obtained, it was
shown that D=11 would be the highest possible dimension because
beyond it, SUSY would require \cite{nahm} spin $>$2 and more
gravitons to keep the boson/fermion degree of freedom balance.  As
mentioned, these are physically inconsistent systems, except of
course for the dull case of uncoupled free fields.  The D=11
theory itself was completely established as well \cite{cremmer}.
Because of its amazing properties, and because of its recent
``post D=10 string" resurgence as as part of the (as yet unknown)
M-theory, it has once again become popular in other contexts.  The
action, of which we only write the bosonic terms here, contains
only three basic fields: the graviton, a 3-form potential
$A_{[{\mu\nu}\alpha ]}$ and its curl, the gauge invariant field
strength $F_{{\mu\nu}\alpha\beta}= \partial_{[\beta}
A_{{\mu\nu}\alpha ]}$ 4-form, together with a single vector-spinor
$\psi_\mu$; schematically,
\begin{equation}
I_{(\mbox{\rm boson})} = \kappa^{-2} \int d^{11}x \:
 \sqrt{-g} \:
\left[ R +  F^2 \right]
 + \kappa
\int d^{11}\! x \:
 \epsilon^{1..11}
A_{123} F_{4..7}F_{8..11} \; . \end{equation} Note the
Chern--Simons term ($\epsilon$ is the Levi-Civita density) and its
explicit $\kappa$-dependence, with $\kappa^2 \sim [L]^9$ here.
This model is unique: there are no different ``$N>1$" versions
possible, there is no ``supermatter" which can serve as its source
and it cannot even (unlike D$<$11 SUGRAs), acquire a nonvanishing
cosmological constant \cite{bautier}.

What are the obstacles to determining the counterterms here?  The
first is that their form is hard to find because of the absence of
a  manifestly SUSY invariant formalism in D=11. Because $D$ is
odd, the simplest place to look is at two-loop order,
\begin{equation}
\Delta I_{\mbox{\rm 2-loop}} \sim \kappa^2 \int d^{11}x
\sum^0_{n=2}
 [R_1D^{20-n}R_n + ...] \; .
\end{equation} This rather symbolic expression is intended to mean
any dimension 20 invariant made from curvatures and derivatives,
plus its SUSY completion if one exists. Clearly, the more D's and
fewer $R$'s, the easier our task. However, just from linearized
SUSY arguments, ``$R^2$" is trivial (because Gauss--Bonnet holds
at any D for the leading quadratic terms, while $R^3$ is likewise
not useful as the leading term of a SUSY invariant.  Thus,
$R^4D^{12}$ is the most likely place to start.

I will summarize here the work of Seminara and myself \cite{sdds}
in which the invariant term has been obtained. The method consists
in realizing that there is a guaranteed linearized SUSY invariant
to be constructed from the action (7) -- the four-point tree level
scattering amplitude -- that is, the sum of all possible 4
particle scatterings, from 4-graviton to 4-fermion. Furthermore,
because linearized SUSY does not change particle number, and
because at tree level bosons and fermions doesn't mix in the sense
that the fermions do not affect the bosonic amplitudes, it
suffices to consider only the bosonic truncation of (7) to get all
4-boson parts.  Before describing the calculation further, we must
state that these amplitudes are nonlocal, because they involve
exchange of intermediate particles, with accompanying propagator
denominators, and they have fewer than the required explicit
derivatives besides. We will have to remove the nonlocality and
insert further derivatives in a way consistent with SUSY for this
construction to be useful; this is indeed possible.  I do not give
details here; suffice it to say that the operative quantities are
(a) the graviton and form propagators, both $\sim p^{-2}$, (b) the
three-point vertices, proportional to $\kappa p^2$, and (c) the
four-point ones $\sim \kappa^2p^2$.  The propagators are
well-known in any gauge; there are three types of 3-point
vertices, namely the cubic Einstein terms, the cubic pure form
Chern--Simons vertex and finally the coupling of the form's stress
tensor to the graviton.  The four-point contact vertices are
needed for gauge-invariance; their local contributions leave net
nonlocal amplitudes; the quartic Einstein term does this for the
4-graviton amplitude, while the quartic $FFhh$ term that comes
from expanding the kinetic term $\sqrt{-g} \:
F_{{\mu\nu}\alpha\beta} F_{\mu '\nu '\alpha '\beta '} g^{\mu\mu
'}... g^{\beta\beta '}$ to ${\cal O} (h^2)$ is needed for
graviton-form ``Compton scattering".

With the above elements, it is easy to see what processes are
possible. The first is 4-graviton scattering, just as in Einstein
theory. Its amplitude is essentially dimension-independent and can
be expressed either in ``string"-form $t_8 t_8 \, RRRR$ (where
$t_8$ is a constant 8-index tensor), as it appears on expanding
the gravity sector of string theory beyond the leading Einstein
part, or in Bel--Robinson form $\sim B^2_{{\mu\nu}\alpha\beta}$,
as it first arises in D=4 SUGRAs. At the ``opposite" end is the
4-form $\sim F^4$ scattering: it involves the CS-vertices joined
with a form propagator plus the $T_{\mu\nu} (F) - T_{{\mu\nu}}
(F)$ graviton-mediated scattering. Next comes a ``graviton
bremsstrahlung" $F^3R$ term in which a single graviton is emitted
by the $T_{\mu\nu} (F) h_{\mu\nu}$ coupling from any arm of the CS
term (the topological CS vertex itself is of course
metric-independent). Finally, most complicated is the $FFRR$
scattering involving all of the vertices. [There is no $FR^3$ term
at tree level, since a single matter particle cannot be created.]
At this stage, our results may be summarized schematically by the
statement that the total bosonic amplitude is of the -- very
symbolic -- form
\begin{equation}
stu \: M_{\makebox{\rm tot}} \sim [R^4 + R^2F^2 + RF^3 + F^4]
 \; .
\end{equation}
 Here we are using momentum space form where ({\it s,t,u}) are
the usual Mandelstam variables quadratic in momentum used to
describe 4-point scatterings in QFT. There is work involved in
obtaining (9), because we must, and can, establish that
multiplying by stu leads to a local quantity after ``spreading"
the derivatives over the four field strengths and that SUSY is not
lost.  Then we must multiply in an {\it stu}-symmetric way by 12
more derivatives to get the right dimension, which is accomplished
by any combination of ($s^6 + t^6 + u^6$) and ({\it stu})$^2$.

At this stage we have accomplished the desired construction of a
4-point linearized SUSY invariant.  The next question is of course
if it actually appears as a two-loop infinite counterterm, {\it
i.e.}, whether its coefficient is non-zero. This has been answered
in the affirmative in \cite{bern}. Using the methods mentioned
earlier, namely using rules relating gravitational to squares of
Yang--Mills amplitudes \cite{kawai}, it was possible to calculate
the infinite part of the 2 loop 4-graviton counterterm, {\it
i.e.}, of the first term in (9), with dimensional regularization
in D=11-2$\epsilon$ dimensions.  The result is
\begin{equation}
\Delta I^{\mbox{\rm 4 grav}}_{\mbox{\rm 2 loop}} = \kappa^2 \:
A/\epsilon[438 (s^6 + t^6 +u^6 ) - 53 (stu)^2 ] stu \:
M^{\mbox{\rm 4 grav}}_{\mbox{\rm tree}} \end{equation} where $stu
\: M^{4g}_{\mbox{tree}} \sim R^4$ and $A$ is a known constant.
Barring the infinite unlikelihood that there exist hidden
(super)symmetries that are respected by this, but by no higher
loop candidate terms, we may conclude that D=11 SUGRA is also
nonrenormalizable.

\section{Conclusions}

We have seen in a systematic way that there is no local, unitary
(ghost/tachyon-free) quantum field theory whose action reduces to
QGR or classical GR that is also free of infinities; the latter
are almost certainly there at every order, requiring an infinite
number of input parameters to define these theories. The
conclusion includes all possible SUGRA models, {\it i.e.}, from
D=4 through D=11, as well. Although the presence of new
counterterms at all loop orders (or at an infinite set of them)
cannot reasonably be rigorously demonstrable, the fact that the
ones we did see appeared at lowest permitted order (so that no
``hidden" invariances prevented them), is quite convincing
evidence.  The calculations were all within the perturbative
framework in which we know how to define and carry out the
problem, and may not -- perhaps -- apply to some putative future
nonperturbative formulation. [Remember though that perturbation
loop expansion is responsible for all the incredible quantitative
successes of the standard model!]

It seems safe to conclude then, that local QFT's of gravity must
be regarded as only effective low energy (compared to Planck
scale!) approximations to deeper theories of extended objects,
such as strings or M-theory, of which they are the pointlike
limits. The D=11 invariant given here could, from this point of
view, also be considered as a first local finite correction that
emerges from M-theory.

\section{Acknowledgements}

I thank my collaborator D.\ Seminara for helpful discussions.
This work was supported by the National Science Foundation under
grant PHY99-73935.

\end{document}